\documentclass[preprint2]{aastex}
\usepackage{graphicx}

\newcommand{\beq}{\begin{equation}}
\newcommand{\eeq}{\end{equation}}
\newcommand{\bey}{\begin{eqnarray}}
\newcommand{\eey}{\end{eqnarray}}

\newcommand{\grad}{{\bf \nabla}}

\newcommand{\mut}{{\tilde\mu}}
\slugcomment{}
\shorttitle{Refining MOND/TeVeS Lagrangian}
\shortauthors{Zhao \& Famaey}

\begin{document}
\title{Refining MOND interpolating function and TeVeS Lagrangian}
\author{H.S. Zhao\altaffilmark{1}}
\affil{SUPA, University of St Andrews, KY16 9SS, Fife, UK}
\email{hz4@st-andrews.ac.uk}
\and

\author{B. Famaey\altaffilmark{2}}
\affil{Institut d'Astronomie et d'Astrophysique, Universit\'e
Libre de  Bruxelles, CP 226, Boulevard du Triomphe, B-1050 Bruxelles,
Belgium}
\email{bfamaey@ulb.ac.be}
\altaffiltext{1}{PPARC Advanced Fellow}
\altaffiltext{2}{FNRS Scientific Collaborator}

\begin{abstract} The phenomena customly called Dark Matter or Modified Newtonian Dynamics
(MOND) have been argued by Bekenstein (2004) to be 
the consequences of a covariant scalar field, controlled by a free function
(related to the MOND interpolating function $\mut(g/a_0)$)
in its Lagrangian density.  In the context of this relativistic MOND theory
(TeVeS), 
we examine critically the interpolating function in the transition zone between weak and strong gravity.  
Bekenstein's toy model produces too gradually varying $\mut$ and fits  
rotation curves less well than the standard MOND interpolating function $\mut(x)=x/\sqrt{1+x^2}$.  However, the latter 
varies too sharply and implies an implausible  
external field effect (EFE).  These constraints on opposite sides have not
yet excluded TeVeS, but made the zone of acceptable interpolating functions
narrower. An acceptable ``toy" Lagrangian density function with simple analytical properties is singled out for future studies 
of TeVeS in galaxies. We also suggest how to extend the model to solar
system dynamics and cosmology, and compare with strong lensing data (see
also astro-ph/0509590). \end{abstract} 

\keywords{gravitation - dark matter - galaxy kinematics and dynamics}

\section{Introduction}
On galaxy scales, dark matter generally dominates over baryons (stars plus
gas) at large radii.  At intermediate radii in a galaxy where dark matter and 
baryons overlap with comparable amounts, the two mass profiles are not uncorrelated (McGaugh 2005).
The correlation between the Newtonian gravity of the baryons ${\bf g}_N$
and the overall gravity ${\bf g}$ (baryons plus dark matter) can be loosely
parameterized by the Milgrom's (1983) empirical relation
\begin{equation}\label{mueq}
        ~{\mut}(g/a_0) \, {\bf g} = {\bf g}_{{\rm N}},
\end{equation}
where the interpolating function $\mut(x)$ is a function which runs smoothly from
$\mut(x)=x$ at $x\ll 1$ to $\mut(x)=1$ at $x\gg 1$ with a dividing gravity scale $a_0 \sim 10^{-8} {\rm cm}\,{\rm s}^{-2} \sim cH_0/6$
at the transition.  This simple correlation was taken as the basis for the MOND theory (or more precisely the aquadratic Lagrangian theory) 
by Bekenstein \& Milgrom (1984, hereafter BM84), where one modifies the Newtonian gravity of a baryonic galaxy to 
eliminate the need for dark matter.

%even though it could as well be an
%empirical relation between dark matter and baryons.
%even though it remains
%challenging for N-body and hydrodynamical simulations to undertand it in the dark matter picture.  
%On the other hand, the MOND picture is not without problems on sub-galactic scale (Zhao 2005) and on galaxy cluster scale (Sanders 2003).

Recently interests on the subject of MOND have been further stimulated since Bekenstein (2004, hereafter B04) provided 
a Lorentz-covariant theory (dubbed TeVeS), which passes standard tests to check General Relativity, 
and allows for rigourous modeling of Hubble expansion and gravitational
lensing (e.g. Zhao et al. 2005). 
In TeVeS the MONDian behaviour originates from a scalar field, the dynamics
of which is controlled by a Lagrangian density involving a free function that yields the expected dynamics in the low-acceleration limit
(although BM84 theory is not precisely recovered). This freedom of the
Lagrangian density,
that echoes the freedom in the choice of the interpolating function $\mut$
in MOND, means that every choice of the free function defines a distinct
theory. As this class of theories do not at present derive from any basic
principle and are purely phenomenological, the only constraints on the free
function must come from phenomenological grounds. A refinement of the
function studied by B04 as a toy model is surely needed. In this letter, we
differentiate popular choices of the MOND $\mut$ function by fitting a benchmark rotation curve, and argue that many
of those functions are likely unphysical in the TeVeS
context. We then propose a new free function for TeVeS in the
domain relevant for galaxies, with a possible extension to solar system
dynamics and cosmology. 

\subsection{Warming up to TeVeS}
It is a tensor-vector-scalar Lorentz-covariant
field theory, where the tensor is the
Einstein metric $g_{\alpha \beta}$ out of which is built the usual Einstein-Hilbert action, 
$U_\alpha$ is a dynamical normalized vector field, and $\phi$ a dynamical scalar field. The action is the sum of the Einstein-Hilbert action for the tensor $g_{\alpha \beta}$, the matter action, the action of the vector field $U_\alpha$, and the action of the scalar field $\phi$. 
Einstein-like equations are obtained for each of these fields by varying the action w.r.t. each of them.
The action of the scalar field $\phi$ involves a dimensionless parameter $k$ (of the order of a few percents), 
a length scale parameter $l$ ($\sim {\sqrt{3k} c^2 \over 4\pi a_0}$), and a free dimensionless function linking $k l^2 |\grad \phi|^2 \propto y$ 
with the auxiliary nondynamical scalar field $\mu$.

More relevant to us, the physical metric in TeVeS near a quasi-static
galaxy or the solar system is identical to that of General Relativity, with a potential
\beq\label{twopart}
\Phi = \Xi \Phi_N + \phi
\eeq
where $\Xi \simeq 1$. This means that the scalar field $\phi$ plays the role of the dark matter gravitational potential. It is related to the Newtonian potential $\Phi_N$ (generated by the baryonic density $\rho$) through the equation (similar to the field equation for the full $\Phi$ in BM84)
\beq\label{poisson}
\grad . [ \mu_s \grad \phi] = \grad^2 \Phi_N = 4 \pi G \rho, 
\eeq
where $\mu_s$ is a function of the scalar field strength $g_s=|\grad \phi|$.  It is related the $\mu$ function of B04
and the interpolating function $\mut$ of MOND by
\beq\label{mus}
\mu_s \equiv {\mu \over k'} = \frac{\mut}{1-\mut}, 
 ~k' \equiv {k \over 4 \pi}. 
\eeq
In the intermediate to deep-MOND regime, the toy free function in the scalar action of B04 gives rise to the following interpolating function:
\begin{equation}\label{btoy}
\mut(x) = {\sqrt{1 + 4x} -1\over\sqrt{1 + 4x} +1}.
\end{equation}

\section{Rotation Curves}
Milgrom's formula (\ref{mueq}) is a good approximation of the BM84 and TeVeS
theories since for most disk galaxies the curl field is negligible or
exactly zero when solving Eq.~(\ref{poisson}) (Brada \& Milgrom 1995).  It
provides a plausible unified picture for the phenomenology of individual
galaxies, which is more challenging to understand in the dark matter
framework. However, while the precise functional form of the interpolating
function is not necessary to make many fundamental predictions (see Sanders
\& McGaugh 2002), it is nevertheless important in order to fit the rotation
curves of galaxies. Indeed one of the most striking successes of MOND is
the ability of Milgrom's formula to fit the rotation curves of a wide range
of galaxies with virtually the same value of $a_0=1.2 \times 10^{-8} {\rm cm}\,{\rm s}^{-2}$, and the same ``standard'' interpolating function: 
\beq\label{standard}
\mut(x)={x \over \sqrt{1+x^2}}.
\eeq
This ``standard'' interpolating function was originally put in by hand, and
does not derive from any physical principle. However, with the number of
galaxies with good data increasing, the freedom of this function should be
restrained by the observations. Famaey \& Binney (2005, hereafter FB05)
have e.g. found that the ``simple'' interpolating function,
\beq\label{simple} 
\mut(x)={x\over 1+x},
\eeq
gives a better fit to the terminal velocity curve of the Milky Way than
Eq.~(\ref{standard}), while yielding an extremely good fit to the rotation curve of the
standard external galaxy NGC3198 (cf. Fig.~1).  In comparison the toy model
in B04 gives rise to Eq.~(\ref{btoy}) in spherical symmetry in the intermediate to
deep-MOND regime.  This interpolating function triggers too slow a
transition from the MONDian to the Newtonian regime in the benchmark
rotation curve of NGC3198 (cf. Fig.~1); the same conclusion is reached for
the terminal velocity curve of the Milky Way (FB05).  

In short, from the analysis of rotation curves, Eq.~(\ref{standard}) and Eq.~(\ref{simple}) are
preferred over Eq.~(\ref{btoy}). FB05 noted that Eq.~(\ref{simple}) is preferred over Eq.~(\ref{standard}) to
fit the TVC of the Milky Way. They also derived the best MONDian fit (in
the strong and intermediate regimes only) of the Milky Way, and found the
interpolating function to transition smoothly from Eq.~(\ref{simple}) at $x \leq 1$ to
Eq.~(\ref{standard}) at $x \geq 10$. Of course, it is not  obvious that all
the other
galaxies will give the same answer as the Milky Way and NGC3198. The
observational constraints presented in this section should thus be
considered as an indication more than a rigorous constraint.

\section{The external field effect}
In TeVeS the potential consists of two parts (Eq.~\ref{twopart}): the Newtonian potential, and the scalar field which satisfies the BM84 formulation (Eq.~\ref{poisson}). 
One of the key features of the BM84 theory is the existence of 
the dilation effect of an external field, which is why MOND does not
satisfy the strong equivalence principle.  Consider the perturbation generated by a low-mass $m$ body inside 
a dominating uniform external scalar field strength $g_s^e \hat{z}$, Eq.~(\ref{poisson}) for $\phi$ becomes 
\beq
\grad . [\mu_s \grad \phi] \approx \mu_s^e \left(\Delta_1 \partial_z^2 + \partial_x^2 + \partial_y^2 \right)\phi = 4\pi G m \delta(r),
\eeq
which differs from the Newtonian Poisson equation by the dilation factor
$\Delta_1$ given by (BM84; Zhao 2005; Zhao \& Tian 2005)
\beq
\Delta_1 = 1 + {d \ln \mu_s^e \over d \ln g_s^e}, \qquad \mu_s^e \equiv \mu_s(g_s^e).
\eeq
As in BM84, the solution is an equal potential surface given by
\beq\label{pert}
\phi (x,y,z) = - g_s^e z - {G  m \over  \mu_s^e \tilde{r}},~\tilde{r}=\sqrt{z^2 + \Delta_1\left(y^2+x^2\right)}, 
\eeq
where $m /\mu_s^e$ is the effective mass of the satellite 
and $\tilde{r}$ is the effective distance from the centre of the satellite.
Hence the perturbation scalar field is stretched in the $z$ direction by a factor $1/\sqrt{\Delta_1}$. 
To exclude an imaginary dilation, a theory should have 
\beq
\Delta_1  = {d \ln \mu_s g_s \over d \ln g_s} > 0.
\eeq
This requires models where $g_s$ is an increasing function of $g_N=\mu_s g_s$.
The factor $\Delta_1$ also determines the shape of the Roche Lobe of a rotating satellite, 
which have a middle-to-long axis ratio $\sqrt{2 \over 3\Delta_1}$ (Zhao
2005; Zhao \& Tian 2005).
In fact the stretching factor $\Delta_1^{-1/2}$ enters almost all processes 
involving satellites (e.g. Brada \& Milgrom 2000).
For these reasons we consider models with a negative $\Delta_1$ unphysical. 
In the original BM84 theory, this is not a problem because $1 < \Delta_1 <2$ thanks 
to a monotonically increasing $\mut(g)$ (not $\mu_s(g_s)$ as we are concerned with), so the
external field effect is a mild curiosity.  
This, however, is not the case if the MOND effect is created by a scalar field.
From Eq.~(\ref{mus}), we have that models with the standard $\mut$ function (Eq.~\ref{standard}) yield a $g_s$ which increases with $g_N$ 
to some point, and then starts decreasing in the intermediate regime (see
Fig.~2).  At the same scalar field strength $g_s$, there are two different
(spherical) Newtonian 
gravity strength, i.e., the scalar function $\mu_s(g_s)$ becomes multi-valued for the same $g_s$, hence
is ill-defined.\footnote{
The $\mu_s(s)$ is given by two root branchs of an essentially 4-th order polynomial equation $s={\mu_s \over  (1+\mu_s) \sqrt{1+2\mu_s}}$, 
and is a very lengthy multi-valued function.}
This is a general feature of any sharply increasing MOND $\mut$ function
(e.g., the ``exponential" function $\mut=1-\exp(-x)$, and the best fit of
FB05).  This type of behaviour is undesirable in a physical model for the
scalar field. On the other hand the simple function (Eq.~\ref{simple}) and B04 toy model (Eq.~\ref{btoy}) give physical monotonic $\mu_s(g_s)$, hence positive $\Delta_1$.  

\section{New free function}
As a toy model B04 chose for the $\mu$-function (see Eq.~(\ref{mus}) above) the
implicit, discontinuous formulation: 
\beq\label{yf}
{y(\mu) \over 3 k'^2} \equiv 
{h^{\alpha \beta} \phi_{,\alpha} \phi_{,\beta} \over a_0^2}
= \frac{\mu^2}{k'^2} \frac{(\mu -2)^2}{4(1-\mu)}, 
\eeq
where notations are as in B04.  
Here cosmology ($\mu>2$) and galaxies ($0<\mu<1$) are completely detached
from each other, while the fit to galactic rotation curves is poor.

Our aim here is to propose a new 
{\it explicit, monotonic and continuous} 
interpolating function $\mu_s(s)={\mu \over k'}$.
A simple choice could be
\beq\label{toys}
s \equiv {g_s \over a_0} \equiv \sqrt{|y| \over 3k'^2} = {\mu \over (k' + \mu) (1-\mu)^n}, ~n=0,1.
\eeq
Here $y>0$ and $0<\mu<1$ corresponds to quasi-static systems as in B04, and
$y<0$ to cosmology. This toy function is consistent with rotation curves,
since it reduces to the simple interpolating function Eq.~(\ref{simple}) in
the range for galaxies, i.e., $\mu = k'\mu_s \sim (0 - 10)k' \ll 1$, where
we have the explicit relations
\beq
\mu_s \approx {g_s \over a_0 - g_s}, \qquad 
s \approx \frac{\mu_s}{1 + \mu_s},
\eeq
which are ready to be fed into a solver for Eq.~(\ref{poisson}).  

\subsection{Solar system}
Our toy function also recovers solar system dynamics at a level 
similar to B04, if not better.  For example, in $n=1$ case 
(the default case), the scalar field goes to 
infinity when $\mu$ approaches one. This new ``toy" function is easier to use than B04 because it allows for an analytic formulation for the scalar field interpolating function 
$\mu_s(s) \equiv {\mu \over k'} = \frac{\mut}{1-\mut}$ in both strong and weak gravity, namely
\beq\label{propmus}
k' \mu_s(s) = {1 - k' \over 2 } + { -1 + \sqrt{ \left[(1+k')s + 1\right]^2 - 4s } \over 2 s}.
\eeq
The dilation factor $\Delta_1 = 1+{d\ln \mu_s \over d\ln s} = {2+ (1-k') \mu_s \over 1+ k' \mu_s^2} \sim (1-4)$ is in the plausible range for $k'=0.03-1$.
As we can see from Fig.~\ref{newf}, the new function yields a monotically increasing scalar field strength
$g_s$, and the correction for the solar system dynamics is less than in B04
toy model (when adopting $k=0.01$, the variation from Sun to Saturn is an order
smaller), which may or may not be relevant to the Pioneer Anomaly.

\subsection{Cosmology}
Although the B04 discontinuous cosmological branch of the 
function $y(\mu)$ 
could as well be appended to our function in the range $\mu > 2$,
it is more desirable to be in a universe where 
the Lagrangian density has a smooth transition between 
weakly gravitating quasi-static systems 
($y \propto (\grad \phi)^2/a_0^2>0$) and cosmology 
($y \propto -2(\partial_t \phi)^2<0$).  A possible way, 
as done here (Eq.~\ref{toys}), is to copy our function
in $y>0$ regime into the $y<0$ regime by a simple mirror-imaging,  
(cf. Fig.~\ref{cos}). 
This means that the outer parts of galaxies would connect
smoothly into the cosmological expansion as $y$ passes from $0+$ to
$0-$.  While a mirror-image with fine tuning is attractive, 
it is not a necessary condition: any positive continuous function
$\mu(y)$ going through, e.g., 
$\mu(0-)=\mu(0+) \sim 0$ is worth exploring. 
The next question is whether this kind of cosmology 
can produce realistic Hubble expansion and the Cosmic Microwave Background.

\section{Conclusion}
Part of the amazing successes of the non-relativistic version of MOND 
in explaining galaxy dynamics is due to its ``standard" interpolating function $\mut = x/\sqrt{1+x^2}$.
When exploring a range of other empirical functions in the context of TeVeS, we see that the fit 
to rotation curves becomes poorer for more gradual $\mut$ function, whilst an external field
effect with imaginary dilation happens for more rapid changes of the $\mut$ function.  
These two independent constraints from opposite sides suggest a fairly 
narrow range of TeVeS free functions.  
Among these we propose a simple expression which works for both very weak
and very strong gravity (Eq.~\ref{propmus}), with a possible extension to
cosmology.
Unlike the one in B04, the new function has the nice feature that it links 
quantities of TeVeS and of MOND easily, hence
facilitates future examinations using galaxy dynamics and solar system data.
The explicit simple monotonic interpolating function $\mu_s(g_s/a_0)$ could be easily 
fed into a numerical solver for Eq.~(\ref{poisson}) 
(e.g., as developed by Ciotti et al. 2005), and could allow for the
modelling of realistic galaxy geometries (the curl-field of BM84, neglected here following conventional wisdom, could be put back with a realistic amount).  Combined with a galaxy 
with sensitive kinematical data, this may confirm or falsify our ``toy" function, hence further
establish or squeeze the parameter space of the TeVeS theory. 

As two final remarks, (i) we note that multiple-imaged 
gravitational systems present a challenge to all MOND/TeVeS interpolating functions (cf. Fig.~\ref{lens}).
Zhao et al. (2005) found that elliptical galaxies of comparable luminosity and redshift  
show a large scatter in their Einstein ring sizes.  Among the previously proposed $\mu$ functions
the B04 toy function is the most effective in lensing, but none of the functions 
seems to fit all lenses in the point lens case (shaded zone); the fit is poorer with realistic  
lens mass profile (cf. Fig. 17 of Zhao et al. 2005).
(ii) We also note that the dark matter potential is fundamentally different from the scalar field; 
although the two are sometimes degenerate in fitting rotation curves, there is no equivalent
of EFE in dark matter, hence the dark matter potential enjoys more freedom.

\clearpage

\begin{figure}
\epsscale{2.8}
\plottwo{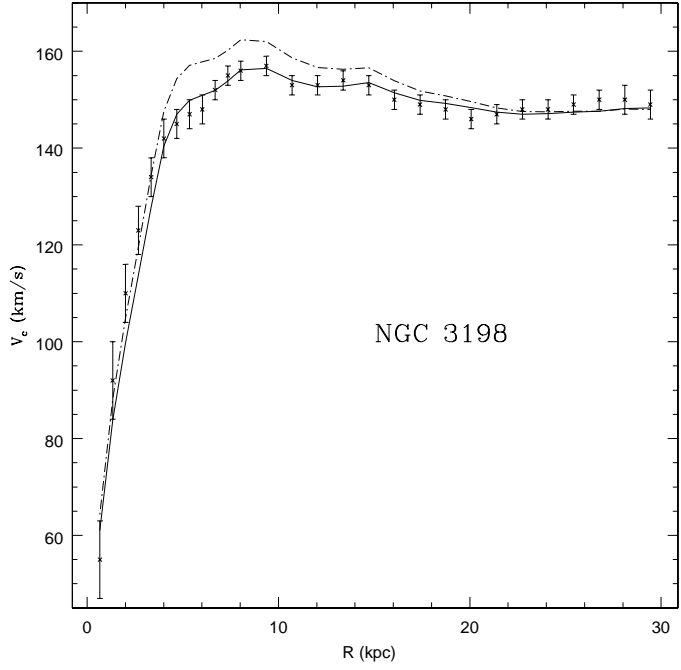}{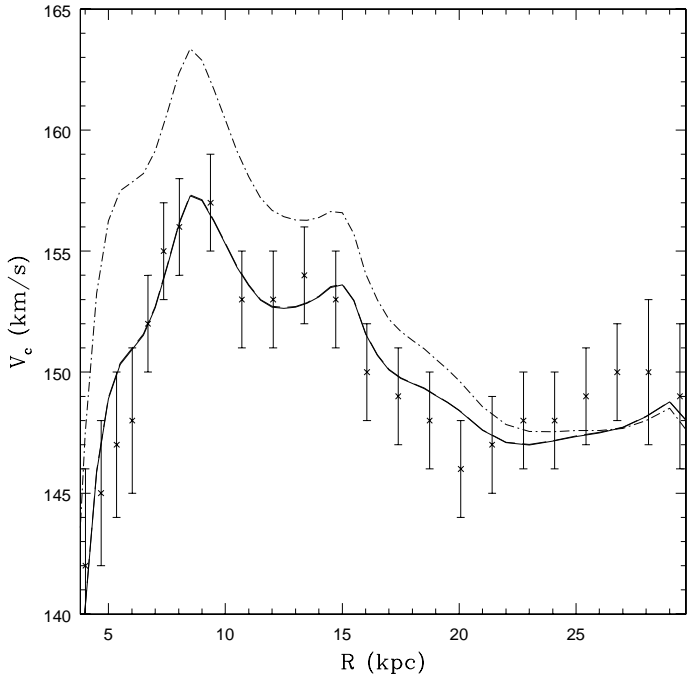}
\caption{Shows the fit to the rotation curve of the ``benchmark" galaxy NGC3198 
using different $\mut$-functions: ``simple" (Eq. (8), solid) and Bekenstein ``toy" functions (Eq. (6), dashed).\label{N3198}}
\end{figure}

\clearpage

\begin{figure}
\epsscale{2.0}
\plotone{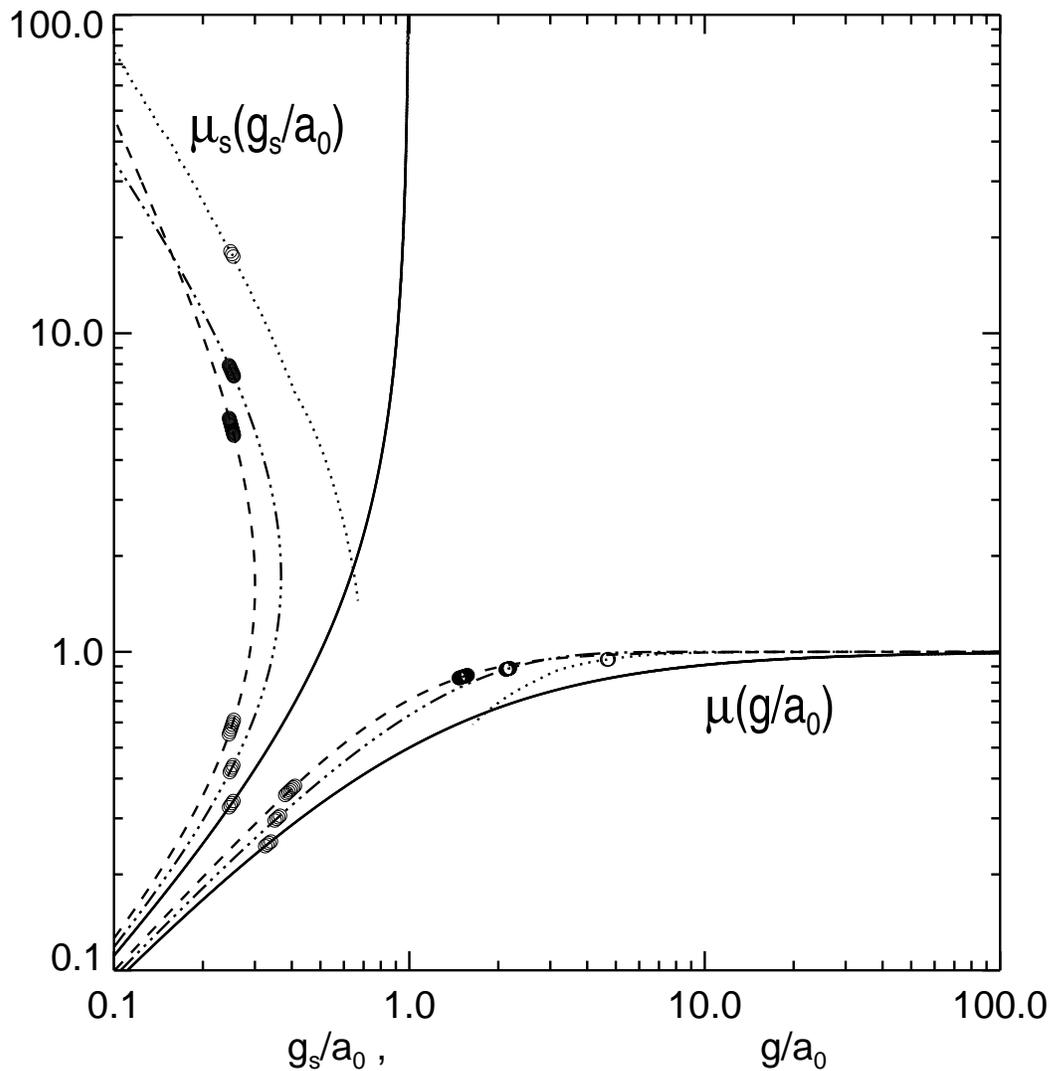}
\caption{shows $x=g/a_0$ vs. $\mut(x)$ (lower right)
for popular choices: ``standard" (Eq. 7, thick dashed), 
``exponential" (dashed-dotted), ``Milky Way" (fit of FB05, dotted) and ``simple" (Eq. 8, solid) 
The corresponding scalar field $s=g_s/a_0$ vs. its modification function $\mu_s(s)$ is also shown (upper left).
Circles mark where $s \approx 0.25$.  Note popular $\mut(x)$ functions often lead to multi-valued
$\mu_s(s)$ curves except for ``simple".
\label{scalarmu}}
\end{figure}

\clearpage

\begin{figure}
\epsscale{2.0}
\plotone{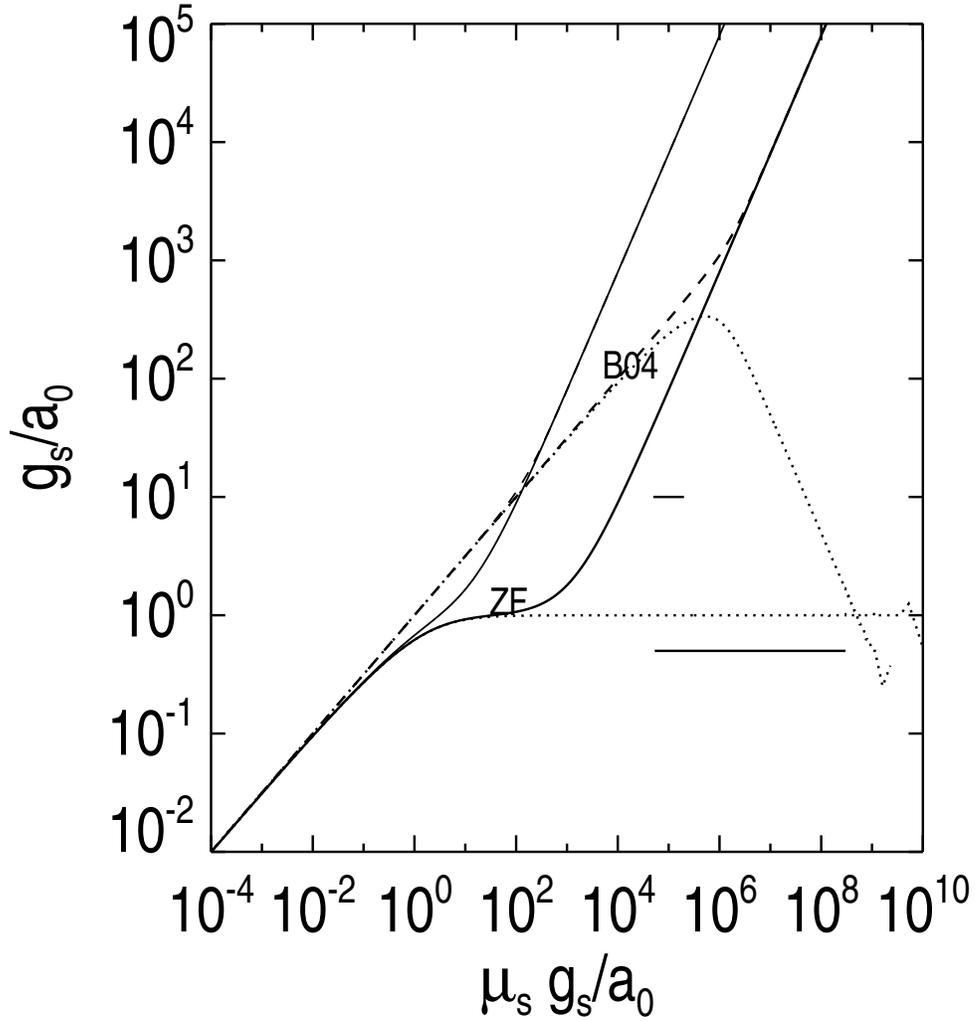}
\caption{Compares our proposed (solid, marked ZF) model with the B04 model (dashed)
in the parameter space of the rescaled scalar field strength $s=g_s/a_0$ vs. Newtonian field strength $g_N/a_0 \equiv s\mu_s(s)$
in cases of $k=4\pi k'=1$ (thick) and $k=0.01$ (thin).
Also overplotted is the deviation from the $\propto
r^{-2}$ force in the solar system (flatter dotted line for our model, other
dotted line for B04). This should be compared with constraints from planets
and the measurement of the Pioneer Anomaly (long and short horizontal thick
lines).
\label{newf}}
\end{figure}

\clearpage

\begin{figure}
\epsscale{2.5}
\plottwo{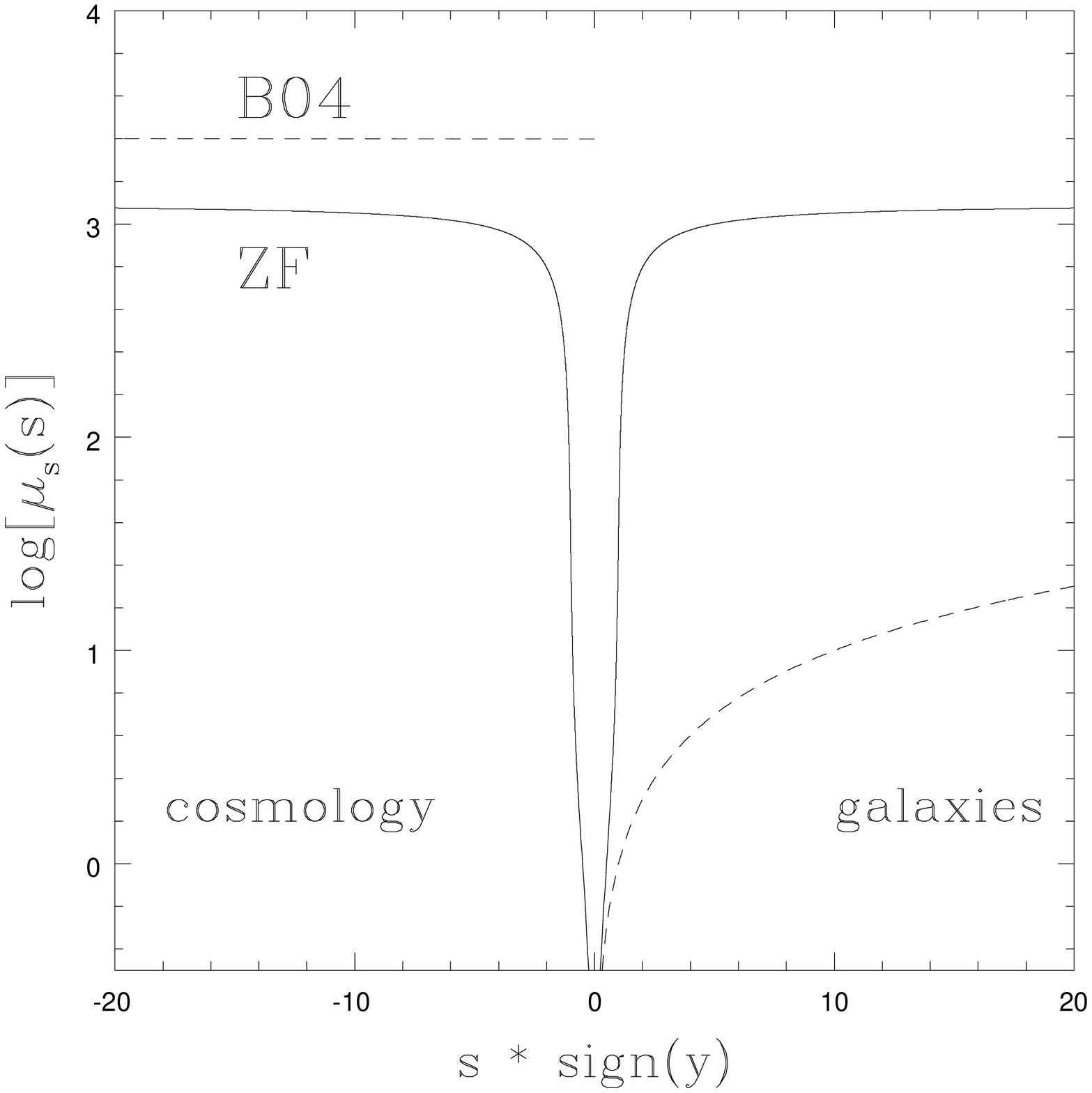}{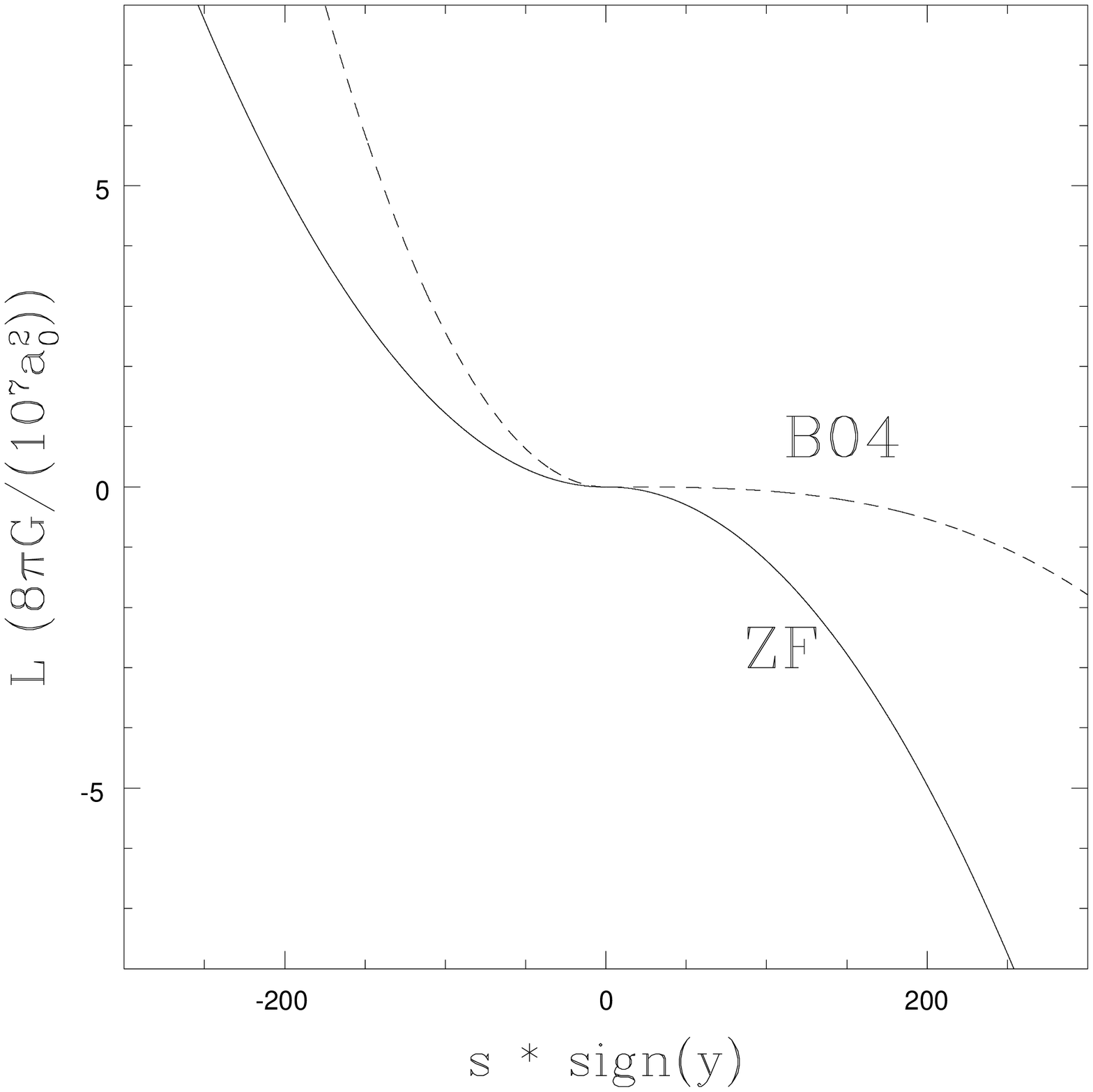}
\caption{The upper panel presents our newly proposed
  function $\mu_s(s)$ (Eq.(16), solid line), compared to B04 toy
  function (dashed). We also present a possible extension to the
  cosmological regime $y<0$ by mirror-imaging (this kind of extension is
  far from being unique). 
  The lower panel displays the above extention of our new Lagrangian density $L$ from the weak gravity $y=0+$ into cosmology $y<0$.
B04 Lagrangian is shown as dashed line for comparison.  Note $y \propto (\grad \phi)^2/a_0^2$ is large and 
positive for the solar system, and $y \propto -(\partial_t \phi)^2$ is negative for cosmology.
\label{cos}}
\end{figure}

\clearpage

\begin{figure}
\epsscale{2.0}
\plotone{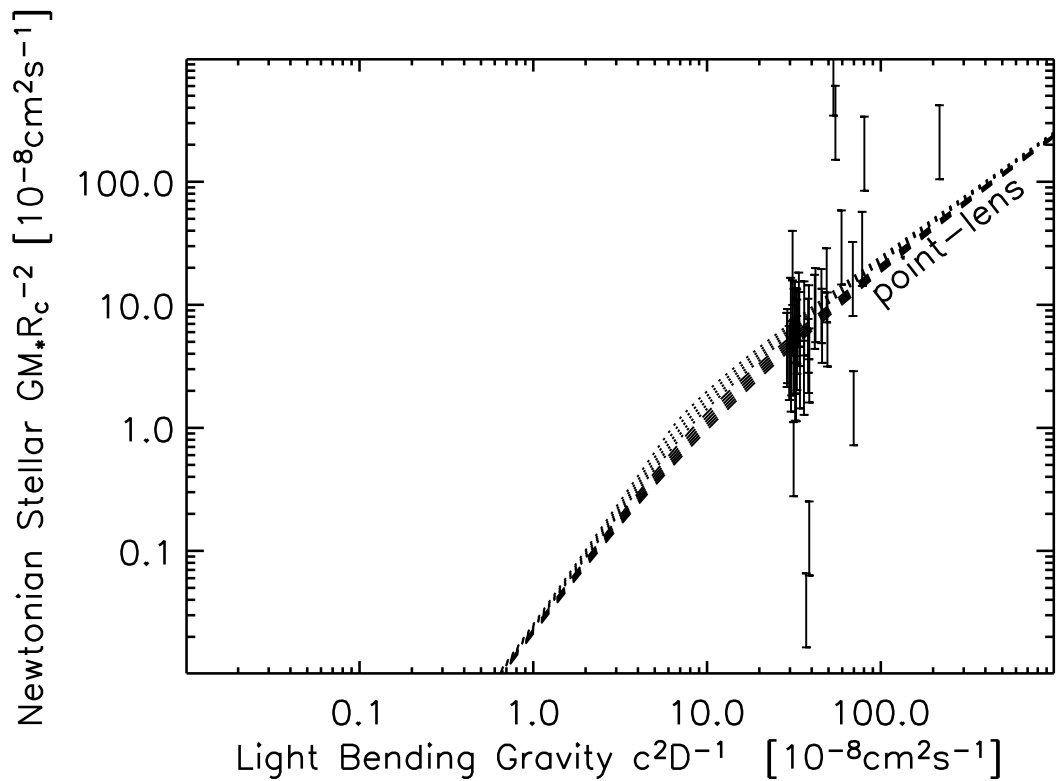}
\caption{Error bars show the stellar Newtonian gravity vs. the light-bending gravity
  for CASTELS sample of galaxy-quasar lenses (Zhao et al. 2005), where we allow for a factor of two uncertainty
  with the stellar mass-to-light ratio.  Predictions are made for point lenses
  with a $\mu$ function in between the B04 toy model (lower boundary of the hatched zone) 
  and a $\mu$ with a sudden transition (upper boundary); models in the upper hatched zone imply unphysical scalar field.
  Galaxies would require more mass to bend the light than the point-lens predictions here.
\label{lens}}
\end{figure}
%\clearpage
\end{document}